\documentclass[11pt]{article}
\usepackage{graphicx}  
\usepackage{a4}
\usepackage{amssymb,latexsym}
\usepackage[mathscr]{eucal}
\usepackage{citesort}
\usepackage{url}

\newcommand{\sMP}{standard Majumdar--Papapetrou\ }

\newcommand{\doc}{{\cal D}_{oc}}
\newcommand{\pdoc}{\partial\doc}

\newcommand{\Uzero}{U_0}

\newcommand{\threeg}{{}g}
\newcommand{\twog}{{}^2g}
\newcommand{\fourg}{{}^4g }

\newcommand{\Sext}{\hyp_{{\mbox{\scriptsize\rm ext}}}}

\newcommand{\zh} {\mathring{h}} 

\newcommand{\hSigma}{\hyp}
\newcommand{\hh}{g}
\newcommand{\hB}{B}
\newcommand{\hM}{Q_B}
\newcommand{\hE}E
\newcommand{\hR}R
\newcommand{\hm}m
\newcommand{\hQ}{Q_E}

\newcommand{\FS}       
                  {F}

\newcommand{\HS} 
       {H_{\mbox{\scriptsize volume}}}

{\ptc{this should be removed in the oberwolfach version}}%

\newcommand{\zA}{\mathring{A}}%
\newcommand{\zZ}{\mathring{P}}
\newcommand{\zT}{\mathring{T}}
\newcommand{\zW}{\mathring{W}}
\newcommand{\zS}{\mathring{S}}
\newcommand{\eeal}[1]{\label{#1}\end{eqnarray}}
\newcommand{\C}{{\mathbb C}}
\newcommand{\bed}{\begin{deqarr}}
\newcommand{\eed}{\end{deqarr}}
\newcommand{\bedl}[1]{\begin{deqarr}\label{#1}}
\newcommand{\eedl}[2]{\arrlabel{#1}\label{#2}\end{deqarr}}

\newcommand{\mcK}{{\mycal K}}

\newcommand{\bel}[1]{\begin{equation}\label{#1}}
\newcommand{\bea}{\begin{eqnarray}}
\newcommand{\bean}{\begin{eqnarray}\nonumber}
\newcommand{\beal}[1]{\begin{eqnarray}\label{#1}}
\newcommand{\eea}{\end{eqnarray}}

\newcommand{\Eq}[1]{Equation~\eq{#1}}

\newcommand{\qed}{\hfill $\Box$ \medskip}
\newcommand{\proof}{\noindent {\sc Proof:\ }}
\newcommand{\be}{\begin{equation}}
\newcommand{\eeq}{\end{equation}}
\newcommand{\ee}{\end{equation}}
\newcommand{\beqa}{\begin{eqnarray}}
\newcommand{\eeqa}{\end{eqnarray}}
\newcommand{\beqan}{\begin{eqnarray*}}
\newcommand{\eeqan}{\end{eqnarray*}}
\newcommand{\ba}{\begin{array}}
\newcommand{\ea}{\end{array}}

\newcommand{\hyp}{\mycal S}

\newcommand{\mcM}{{\mycal M}}

\newtheorem{Theorem} {\sc  Theorem\rm} [section]
\newtheorem{Corollary} [Theorem] {\sc  Corollary\rm}

\newtheorem{Proposition} [Theorem] {\sc  Proposition\rm}

\newtheorem{Remark}[Theorem]{\sc  Remark\rm}

\DeclareFontFamily{OT1}{rsfs}{}
\DeclareFontShape{OT1}{rsfs}{m}{n}{ <-7> rsfs5 <7-10> rsfs7 <10-> rsfs10}{}
\DeclareMathAlphabet{\mycal}{OT1}{rsfs}{m}{n}

{\catcode `\@=11 \global\let\AddToReset=\@addtoreset}
\AddToReset{equation}{section}

\newcounter{mnotecount}[section]

\renewcommand{\themnotecount}{\thesection.\arabic{mnotecount}}

\newcommand{\mnote}[1]
{\protect{\stepcounter{mnotecount}}$^{\mbox{\footnotesize
$
\bullet$\themnotecount}}$ \marginpar{
\raggedright\tiny\em
$\!\!\!\!\!\!\,\bullet$\themnotecount: #1} }

\newcommand{\warn}[1]
{\protect{\stepcounter{mnotecount}}$^{\mbox{\footnotesize
$
\bullet$\themnotecount}}$ \marginpar{
\raggedright\tiny\em
$\!\!\!\!\!\!\,\bullet$\themnotecount: {\bf Warning:} #1} }

\newcommand{\R}{\mathbb R}

\newcommand{\eq}[1]{(\ref{#1})}
\newcommand{\ga}{{\gamma}}
\newcommand{\Mext}{\mcM_\ext}
\newcommand{\ext}{\mathrm{ext}}

\newcommand{\trg}{{\mathrm{tr}_g}}
\newcommand{\divE}{\mathrm{div}(E)}
\newcommand{\divB}{\mathrm{div}(B)}

\newcommand{\whhyp}{\,\,\,\widehat{\!\!\!{\hyp}}}

\newcommand{\hgamma}{\widehat\gamma}
\newcommand{\hgamm}{\hgamma}
\newcommand{\bfxi}{{\bfx}_i}
\newcommand{\bfx}{{x}}
\newcommand{\bfxk}{{\bfx}^k}

\newcommand{\ptc}[1]{\mnote{{\bf ptc:}#1}}

\newcommand{\diag}{\mbox{\rm diag}}

\newcommand{\mcL}{{\mycal L}}

\newcommand{\beqar}{\begin{deqarr}}
\newcommand{\eeqar}{\end{deqarr}}

\newcommand{\beaa}{\begin{eqnarray*}}
\newcommand{\eeaa}{\end{eqnarray*}}

\newcommand{\tr}{\mbox{tr}}

\newcommand{\zhyp}{\mbox{\rm int}\,\hyp}

\begin{document}

\title{On Israel-Wilson-Perj\'es black holes
}

\author{Piotr T.\ Chru\'sciel\thanks{Partially supported by a Polish
Research Committee grant 2 P03B 073 24. E-mail
\protect\url{Piotr.Chrusciel@lmpt.univ-tours.fr}, URL \protect\url{
www.phys.univ-tours.fr/}$\sim$\protect\url{piotr}} \\ LMPT,
F\'ed\'eration Denis Poisson\\
Tours
  \\
  \\
  Harvey S. Reall\thanks{{ E--mail}: \protect\url{Harvey.Reall@nottingham.ac.uk}}
  \\School of Physics and Astronomy, University of Nottingham\\Nottingham NG7 2RD, United Kingdom
\\
\\
  Paul Tod\thanks{{ E--mail}: \protect\url{paul.tod@st-johns.oxford.ac.uk}}
\\
Mathematical Institute and St John's College\\ Oxford}

\maketitle
\begin{abstract}
 We show, under certain conditions, that regular
Israel-Wilson-Perj\'es black holes necessarily belong to the
Majumdar-Papapetrou family.
\end{abstract}


\newcommand{\beq}{\begin{equation}}

\newcommand{\zL}{\mathring{L}}
\newcommand{\zX}{\mathring{Z}}%

\section{Introduction}
\label{SI}

 A classical result in general relativity is the bound on global
charge of regular electro-vacuum space-times, in absolute value,
by the ADM mass, with equality holding if and only if the metric
is, locally, the Israel-Wilson-Perj\'es (IWP)
metric~\cite{Tod,GibbonsHull,Herzlich:mass}, see
Theorem~\ref{Tcharge} below for a precise statement. It is
therefore of interest to classify all non-singular IWP solutions.
A long standing conjecture asserts that those necessarily belong
to the Majumdar-Papapetrou family; for partial results
see~\cite{HartleHawking}, compare~\cite{ChNad}. We prove that this
is indeed the case in electro-vacuum under
supplementary hypotheses.

A key feature which singles out the IWP metrics is the existence of a
``super-covariantly constant'' spinor field
$\psi$~\cite{Tod,GibbonsHull}. Each such spinor leads to a Killing
vector field $X$, which can only be \emph{timelike} or \emph{null}.
Recall that in a regular black hole space-time a Killing vector cannot
be timelike on the event horizon. In IWP space-times
ergoregions\footnote{By \emph{ergoregion} we mean the set where a
``stationary'' Killing vector is spacelike.} do not exist, but null
orbits away from the event horizon could occur in principle. Assuming
that there are no such orbits, we prove that the metric has to be
static.

More precisely, our first main result is the following (see
Section~\ref{Sprel} for definitions):

\begin{Theorem}
  \label{T1n}
  Let $(\mcM,\fourg,F)$ be a solution of the Einstein--Maxwell equations
  with a non-trivial spinor field $\psi$ which is parallel with
  respect to an $F$-modified spinor connection as in
  \eq{sc1}-\eq{sc2}. Suppose that $\mcM$ contains a connected and
  simply connected space-like hypersurface\footnote{We use the
  geometers' convention that a hypersurface with boundary contains its
  boundary as a point set. The signature is $(+,-,-,-)$.} $\hyp$ (with
  boundary), which is the union of an asymptotically flat end and of a
  compact set, such that:
\begin{enumerate}
 \item The Killing vector field $X$ associated with $\psi$ is timelike
 on the interior $\zhyp$ of $\hyp$.
\item The topological boundary $\partial \hyp\equiv
{\hyp}\setminus \zhyp$ of $\hyp$ is a nonempty, two-dimensional,
  topological manifold, with $\fourg_{\mu\nu}X^\mu X^\nu = 0 $ on
  $\partial\hyp$.
  \end{enumerate}
Then, performing a duality rotation of the Maxwell field if necessary,
  there exists a neighborhood of $\hyp$ in $\mcM$ which is
  isometrically diffeomorphic to an open subset of a standard
  Majumdar-Papapetrou space-time.
\end{Theorem}


 For a complete understanding of the problem it is of some
interest to look for a corresponding result without the electro-vacuum
condition.  This is done in Section~\ref{Sadd}.

 Let us give an outline of the proof of Theorem~\ref{T1n};
this also serves as a guide to the structure of this paper.  In
Sections~\ref{Snh} and \ref{Sccs} we examine in detail the space-time
geometry near the event horizon. This allows us to show, in
Section~\ref{Shis}, that horizons correspond to isolated singularities
in space in the usual local coordinate representation of IWP metrics.
This part of our work is purely local, except for the hypothesis of
compactness of cross-sections of the horizons; it follows closely the
calculations in~\cite{CT} and is inspired by the analysis of
supersymmetric black holes in~\cite{Reall:2002bh}. In
Section~\ref{Sas} we analyse the asymptotic behavior of the fields
involved in defining the metric. In Section~\ref{Si} we show that the
local coordinates are global. In Section~\ref{Soa} we establish
staticity of the solutions.

The hypothesis in Theorem~\ref{T1n} that the set $\{g_{\mu\nu}X^\mu
X^\nu = 0 \} $ is a topological manifold of co-dimension one is,
essentially, the condition that all null orbits of the Killing vector
field $X$ lie on the event horizon. This restriction is not needed if
we assume instead that there exists a maximal hypersurface in $\mcM$;
this is the second main result of our work:

\begin{Theorem}
\label{Tmain2} Let $(\mcM, {}\fourg ,F)$ be an electrovacuum
space-time with a super-covariantly constant spinor field
$\psi\not\equiv 0$. Suppose that $\mcM$ contains a simply
connected maximal hypersurface $\hyp$ which is the union of a
compact set with an asymptotically flat region and with a finite
number of ``weakly cylindrical'' regions as in \eq{RH} below. Then
$\hyp$ is totally geodesic and, up to a duality rotation of the
Maxwell field, there exists a neighborhood of $\hyp$ isometrically
diffeomorphic to a subset of a \sMP space-time.
    \end{Theorem}

Theorem~\ref{Tmain2} is proved in Section~\ref{Swmh}.

 The hypotheses of Theorems \ref{T1n} and \ref{Tmain2} do not suffice to obtain more
 information about the size of the set on which the metric is that of
 a \sMP space--time (in the sense defined in Section~\ref{Sprel}). The
 following version of Theorem~\ref{T1n} can be obtained when
 reasonably mild supplementary hypotheses are made:

\begin{Theorem}
  \label{T1.1n0} Let $(\mcM,\fourg,F)$ be a  solution of the  Einstein--Maxwell
  equations containing a connected space-like hypersurface $\hyp$,
  with non-empty topological boundary, which is the union of a finite
  number of asymptotically flat ends and of a compact interior.
  Denote by $\doc\equiv \doc(\Mext)$ the domain of outer
  communications in $(\mcM,g)$ associated with one of the
  asymptotically flat ends of $\hyp$. Let $\psi$ be a non-trivial
  super-covariantly constant spinor field on $\mcM$ and suppose that
  the associated Killing vector field $X$ is timelike on $\doc$.
  Assume morever that
\begin{enumerate}
\item The interior $\zhyp$ of $\hyp$ is a subset of the domain of
outer communications $ \doc$.
\item The topological boundary $\partial \hyp\equiv {\hyp}\setminus
 \zhyp$ of $\hyp$ is a nonempty, two-dimensional, topological manifold
 such that $\partial \hyp = \hyp \cap \pdoc$.
\item   $X$ has  complete orbits in $\doc$.
\item  $(\doc,g|_{\doc})$ is globally hyperbolic.
\end{enumerate}
Then $\doc$ is isometrically diffeomorphic to a domain of outer
communications of a standard extension of a \sMP space--time.
\end{Theorem}

The proof of Theorem \ref{T1.1n0} follows from Theorem~\ref{T1n} by
standard arguments (compare~\cite{Chstatic}) and will be omitted. We
simply note that the properties that $\hyp$ is simply connected and
has only one asymptotically flat end follow from \cite{ChWald}. The
hypothesis of timelikeness of $X$ in $\doc$ can be replaced by that of
existence of a maximal surface with a finite number of asymptotically
flat ends and of weakly cylindrical ends,
invoking~Theorem~\ref{Tmain2}.

\section{Preliminaries}
\label{Sprel}

As we will be using two-index spinor fields it is convenient to
use the signature $(+---)$. The space-time metric will be denoted
by $\fourg$, the associated \emph{Riemannian} metric induced on
space-like hypersurfaces by $g$. In adapted coordinated in which
$\hyp=\{t=0\}$ one thus has $g_{ij}=-\fourg_{ij}$. The symbol
$\nabla$ denotes the space-time covariant derivative operator
associated with $\fourg$.

Recall that the Majumdar--Papapetrou (MP) metrics are, locally, of
the form \cite{Majumdar,Papapetrou:mp} \beqa\label{I.0} & {}\fourg
= u^{-2}dt^2
- u^2(dx^2+dy^2+dz^2)\,, & \\
\label{I.0.1}
&A = u^{-1} dt\,, & \eeqa where $A$ is the Maxwell potential, $F=dA$,
with some nowhere-vanishing, say positive, function
$u$. 
A space--time will be called a \emph{standard MP space--time} if the
coordinates $x^\mu$ of \eq{I.0}--\eq{I.0.1} are global with range
$\R\times(\R^3\setminus\{\vec a_i\})$ for a finite set of points $\vec
a_i\in\R^3$, $i=1,\ldots,I$, and if the function $u$ has the form
\beq
\label{standard}
u=1+\sum_{i=1}^I \frac{m_i}{|\vec x - \vec a_i|} \,,
\eeq
for some
positive constants $m_i$. It has been shown by Hartle and Hawking
\cite{HartleHawking} that every standard MP space--time can be
analytically extended to an electro--vacuum space--time with a
non--empty black hole region, and with a domain of outer communication
which is non--singular in the sense of the theorems proved here. Those
extensions will be called the {\em standard extensions} of the \sMP
space--times.

 A data set $(\Sext,g,K)$ with Maxwell initial data $F=(E,B)$ will be
 called an \emph{asymptotically flat end} if $\hyp _{\mathrm{ext}}$ is
 diffeomorphic to ${\mathbb R}^3$ minus a ball and if the fields
 $(g_{ij},K_{ij})$ satisfy the fall--off conditions ($\rho$ is the
 radial coordinate in ${\mathbb R}^3$)
\begin{equation}
|g_{ij}-\delta _{ij}|+\rho |\partial _\ell g_{ij}|+\cdots
+\rho^k|\partial _{\ell _1\cdots \ell _k}g_{ij}|+\rho|K_{ij}|+\cdots
+\rho^k|\partial _{\ell _1\cdots \ell _{k-1}}K_{ij}|\le
C_{k,\alpha}\rho^{-\alpha }\ ,
\label{falloff}
\end{equation}
for some constants $C_{k,\alpha}$, $\alpha >0$, $k\ge 1$.  We
shall further require that in the local coordinates as above on
$\Sext$ the Maxwell  field $F$ satisfies the fall--off conditions
\begin{equation}
|E _i | +\rho|\partial _\ell E _i|+\cdots +\rho^k|\partial _{\ell _1\cdots
  \ell _k}E _i| +|B _i | +\rho|\partial _\ell B _i|+\cdots +\rho^k|\partial
  _{\ell _1\cdots \ell _k}B _i| \le \hat C_{k,\alpha}\rho^{-\alpha-1 }\ ,
  \label{Efalloff}
\end{equation}
for some constants $\hat C_{k,\alpha}$, $\alpha >0$, $k\ge 0$. We will
always assume $\alpha >1/2$, which makes the ADM mass well defined in
electro-vacuum.\footnote{It follows in any case from \cite[Section
1.3]{Chnohair} or from~\cite{KennefickMurchadha} that in stationary
electro--vacuum space--times there is no loss of generality in
assuming $\alpha= 1$, $k$ -- arbitrary.} A hypersurface will be said
to be \emph{asymptotically flat} if it contains an asymptotically flat
end $\Sext$.

A two-dimensional surface $S\subset \hyp$ will be called
\emph{weakly outer trapped} if it separates $\hyp$ into two
components, and if $\lambda+h_{ab}K^{ab}\le 0$, \emph{or} if
$\lambda-h_{ab}K^{ab}\le 0$, where $h_{ab}$ is the metric induced
on $S$, and where $\lambda$ is the mean curvature of $S$ within
$\hyp$, as measured with respect to a field of unit normals
pointing towards the component of $\hyp\setminus S$ which contains
$\Sext$.

We note a precise version of the charge bound mentioned in the
Introduction (compare~\cite{GHHP,GibbonsHull,Herzlich:mass}):

\begin{Theorem}
\label{Tcharge}
  Let $(\hSigma,\hh,K)$ be a smooth three-dimensional initial
  data set, with $(\hyp,g)$ complete, and with an asymptotically flat end $\hSigma_\ext$ (in the
  sense of Equation \eq{falloff} with $k\ge 4$ and $\alpha>1/2$), and with $\partial\hSigma$
   weakly outer trapped, if not empty. Suppose,  further, that we are given
   on $\hyp$ two smooth vector fields $\hE$ and $\hB$ satisfying
$$
4\pi \rho_B:=D_i B^i \in L^1(\hyp)\;, \quad 4 \pi \rho_E:=D_i E^i
\in L^1(\hyp)\;.
$$
Set
$$4\pi Q^E:= \lim_{R\to\infty}\int_{r=R}E^idS_i\;,\quad 4 \pi Q^B:=
\lim_{R\to\infty}\int_{r=R}B^idS_i\;.
$$
Let $\hR$ be the Ricci scalar of $\hh$
   and assume
 \bel{pec} 0 \le \hR- |K|^2+(\tr K)^2-2\hh(\hE,\hE)- 2\hh(\hB,\hB)=:
 16 \pi\rho_m\in L^1(\hSigma_\ext) \ .
  \ee
   If
 \bel{pc2}
  \rho_E^2+ \rho_B^2+|J|_g^2\le \rho_m^2\;, \ee
  where
  \be
  16 \pi J^i =2 D_j (K^{ij} - \mbox{\rm tr}
K g^{ij}) -4\epsilon^i{}_{k\ell}E^kB^\ell
  \;,
  \ee
  then the ADM mass $\hm$ of $\hSigma_\ext$ satisfies
 \bel{mci}\hm\ge \sqrt{|\vec p|^2+\hQ^2 + \hM^2}\; .
  \ee
  If the equality is attained in \eq{mci} then \eq{pc2} is also an equality, and
  there exists on $\hSigma$ a spinor field satisfying
  \eq{sc1}-\eq{sc2} below. Furthermore, the associated space-time
  metric is, locally, an IWP (\emph{not} necessarily electro-vacuum) metric.
\end{Theorem}

 The conditions \eq{pec} and \eq{pc2} are clearly satisfied in
electro-vacuum.  This theorem justifies the interest in IWP
space-times; its hypotheses further serve as a guiding principle for
the hypotheses of our remaining results in this paper.

We sketch a proof of Theorem~\ref{Tcharge} in Appendix~\ref{Acharge}.

\subsection{IWP metrics}
\label{sSIWP}

Consider a space-time that admits a ``super-covariantly constant
spinor'' given in two-component spinor notation by
$\psi=(\alpha_A,\beta_{A'})$ where the constituent spinors $\alpha_A$
and $\beta_{A'}$ satisfy the coupled system of equations:
\begin{eqnarray}
\nabla_{AA'}\alpha_B+\sqrt{2}\phi_{AB}\beta_{A'}&=&0\label{sc1}\\
\nabla_{AA'}\beta_{B'}-\sqrt{2}\overline{\phi}_{A'B'}\alpha_A&=&0.\label{sc2}
\end{eqnarray}
Here $\phi_{AB}$ is the Maxwell spinor and $\overline{\phi}_{A'B'}$ is
its complex conjugate, related to the Maxwell tensor $F_{ab}$ in the
standard way by
\[F_{ab}=\phi_{AB}\epsilon_{A'B'}+\overline{\phi}_{A'B'}\epsilon_{AB}.\]
(Strictly speaking, the system (\ref{sc1})-(\ref{sc2}) has a
two-complex-dimensional vector space of solutions, but we shall
normalise the solution by choices made later.)

 As is well known (and will be shown in any case below), the metric is
invariant under the flow of the following vector field\footnote{The
vector field $X$ here equals $ K/\sqrt 2$ in~\cite{Tod}. The value of
the normalisation constant $k$ of that last reference is a matter of
convention, and here we take it to be equal to $\sqrt 2$. Note a
factor of $2$ missing in the first term at the right-hand-side of
\cite[(2.13)]{Tod}.}
\bel{Xdef} X=\frac{1}{\sqrt{2}} (\alpha^A\overline{\alpha}^{A'}
+\overline{\beta}^A\beta^{A'})\frac{\partial}{\partial x^{AA'}},
\ee
It is known~\cite{Tod} that near any point at which $X$ is timelike
the metric can locally be written in the IWP form
\beq\label{met1}
ds^2=V\overline{V}(dt+{\bf{\omega}}\cdot d{\bfx })^2-
(V\overline{V})^{-1}d{\bfx }\cdot d{\bfx }
\eeq
To derive (\ref{met1}) from the system (\ref{sc1})-(\ref{sc2}) one
makes a sequence of definitions.  First, introduce
$$V=\alpha_A\overline{\beta}^A\;,$$
note that $V$ is a smooth function on space-time, vanishing only at
those points at which $X$ is null or zero.\footnote{ It follows from
(\ref{sc1})-(\ref{sc2}) that, if $\alpha_A$ and $\beta_{A'}$ both
vanish at a point $p$ then they vanish everywhere, so we may assume
that they have no common zero. Consequently the Killing vector $X$ has
no zeroes: $X$ will be null where $V$ vanishes but is time-like at all
other points. Space-times with $V$ identically zero are
$pp$-waves~\cite{Tod}, which do not concern us here.}  Then introduce
the coordinates ${\bfx }=(x^1,x^2,x^3)$ by solving:
\begin{eqnarray}
\label{xdef1}
dx^1+idx^2&=&\sqrt{2}\alpha_A\beta_{A'}dx^{AA'}\;,
\\
\label{xdef2}
dx^3&=&\frac{1}{\sqrt{2}}(\alpha_A\overline{\alpha}_{A'}
-\overline{\beta}_A\beta_{A'})dx^{AA'}\;,
\end{eqnarray}
where the one-forms on the right are closed by virtue of
(\ref{sc1})-(\ref{sc2}). Using a time coordinate adapted to the
vector field  $X$,
\bel{Xdefo} X=\frac{\partial}{\partial t}\;,
\ee
one writes
\[X_\mu dx^\mu=V\overline{V}(dt+{\bf{\omega}}\cdot d{\bfx })\]
where ${\bf{\omega}}$ is a one-form, to be determined. Here, by
definition
\[^4 g(X,X) =V\overline{V}.\]
{}From (\ref{sc1})-(\ref{sc2}) and definitions made so far one readily
obtains
\begin{eqnarray}
\nabla_{AA'}V&=&-2\phi_{AB}X_{A'}{}^B\label{dV}\\
\nabla_\mu X_\nu&=&\overline{V}\phi_{AB}\epsilon_{A'B'}
+V\overline{\phi}_{A'B'}\epsilon_{AB}.\label{dX}
\end{eqnarray}
This leads to an Einstein-Maxwell space-time with sources
which may be calculated from the following:
 \beal{Max1} &\nabla_\mu F^{\nu\mu} = 4\pi \chi^E  X^\nu\;,
 &
 \\
 \label{Max2}
 &
 \nabla_\mu (*_4 F^{\nu\mu}) = 4\pi \chi^B  X^\nu\;,
 &
 \\
 \label{Tmunu}
 & T_{\mu\nu} = \chi  X_\mu X_\nu + \frac 1 {4\pi}
 \left(F_{\mu\alpha}F_\nu{^\alpha}-\frac 14 { F^{\mu\nu}F_{\mu\nu}}
 \;
{\fourg_{\alpha\beta}}\right)\;,
 &
 \\
 &
 \chi^E+i \chi^B =\chi  V
 &
 \eeal{rel}
 It follows from (\ref{dV}) and (\ref{dX}) that $X$ is a Killing
vector, while $V^{-1}$ is related to $\chi$ as follows
\bel{nonharmeq}
 \displaystyle
 \Delta(V^{-1}):=\left(
  \left(\frac{\partial}{\partial x^1}\right)^2
+\left(\frac{\partial}{\partial x^2}\right)^2
+\left(\frac{\partial}{\partial x^3}\right)^2\right)(V^{-1})=-\frac
 {4\pi \chi}  V, \ee
 We note that it is easy to construct an infinite-dimensional family
of solutions of \eq{nonharmeq} with smooth \emph{real strictly
positive} $V$, and prescribed positive smooth $\chi V^{-1}$. Thus,
there exist many singularity-free non-vacuum solutions with positive
energy density which saturate the bound of
Theorem~\ref{Tcharge}.

More generally, \eq{rel} and the physical requirement that there are no
magnetic currents imposes the non-trivial restriction that $V$ should
be real on the support of $\chi$. This leads to $\Delta \Im (V^{-1})
=0$. Similarly to our main Theorem~\ref{T1n}, we expect that this
forces the space-time to be static (either with or without black
holes), but we have not attempted to prove that.

Finally $\omega$ satisfies the equation
\beq
\rm{curl\;}\omega=i({\overline{V}}^{-1}\nabla V^{-1}-V^{-1}\nabla\overline{V}^{-1}).
\label{curl}
\eeq
Locally, the integrability condition for (\ref{curl}) is satisfied by
virtue of \eq{nonharmeq} since $\chi$ is real. However, there are
global conditions if $\omega$ is to be well-defined and we shall
return to this point. We write $\Omega$ for the set of points in ${\R
}^3$ at which $V^{-1}$ is singular.

{}Except for Section~\ref{Sadd}, from now on we assume that
$\chi\equiv 0$, so that $V^{-1}$ is harmonic in the flat
three-metric:
\bel{harmeq}\Delta(V^{-1})=0,\ee
The source-free Maxwell equations, which we assume hold, are
equivalent to the equation
\[\nabla^{AA'}\phi_{AB}=0.\]
The space-time is electro-vacuum, so that the Ricci spinor is
related to the Maxwell spinor by Einstein's equations which in
this formalism take the form
\beq\label{phi1}
\Phi_{ABA'B'}=2\phi_{AB}\overline{\phi}_{A'B'} \eeq
in units with $G=c=1$.

\section{Local considerations}
\label{SLc}

\subsection{The near horizon geometry of IWP metrics}
\label{Snh}

 Now concentrate on one component of $\Omega$ which by assumption
corresponds to a component of the Killing horizon.

 As in~\cite{Reall:2002bh,CRT,CT} we introduce Gaussian null
 coordinates near a component ${\cal{N}}$ of the event horizon, with
 the signature chosen so that the metric is
\beq
{}\fourg =r\phi du^2-2dudr-2rh_ady^adu-h_{ab}dy^ady^b,
\label{metric1} \eeq
where $a,b$ range over $\{1,2\}.$

In these coordinates, the Killing vector $X$ is $\partial/\partial u$
with norm
\[\fourg (X,X)=r\phi\]
and ${\cal{N}}$ is located at $r=0$, where also $V=0$. Using the
metric (\ref{metric1}) to lower the index on $X$ we find
\[X_\mu dx^\mu =r\phi du-dr-rh_ady^a\]
whence, at ${\cal{N}}$,
\[d(X_\mu dx^\mu)=\phi dr\wedge du-dr\wedge(h_ady^a).\]
However, from (\ref{dX}), we see that $\nabla_\mu X_\nu$ vanishes at
$V=0$. Two things follow from this: the surface gravity
$\kappa=-\partial_{r}(r\phi)$ at $r=0$ vanishes, and so the horizon is
degenerate; and $h_a$ vanishes at $r=0$. It follows that
\[\phi=rA(r,y^b)\;;\;h_a=rH_a(r, y^b)\]
for some function $A$ and covector field $H_a$. We shall often use a circle over a quantity
to indicate its value at $r=0$, e.g. $\zA = A|_{r=0}$.

We shall show the following:
\begin{Proposition}\label{Pgeom}
\begin{itemize}
\item[(i)]
The metric $h_{ab}$ on the spheres ${\cal{S}}=(r=0,\;u=u_0)$ has
constant Gauss curvature $K$.
\item[(ii)]
On ${\cal{N}}$, $A =K>0$, so that $Ah_{ab}|_{r=0}$ is the unit round
metric on $S^2$.
\item[(iii)] The function $V$ satisfies $\partial_r V  = 2Q$ at
$\cal N$ where $Q$ is a complex constant related to $K$ by $K=4|Q|^2$.
\end{itemize}
\end{Proposition}

\proof $V=0$ at $r=0$ implies that
\be
\label{Veq1}
 V = 2r Q + O(r^2).
\ee
for some smooth function $Q$ independent of $r$, with $\zA = 4
|Q|^2$. We start by showing that $Q$ is not identically zero. It
is useful to invoke the \emph{near horizon limit}, as in
\cite{Reall:2002bh}, obtained by introducing new coordinates
$(\hat r, \hat u, \hat y^a)$ defined by the formula
$$ r= \epsilon \hat r\;, \quad  u = \epsilon^{-1} \hat u\;,\quad \hat y^a=y^a\;,
$$
and letting $\epsilon$ go to zero.  Assume, for contradiction, that
$Q$ vanishes identically. By \eq{dV} the Maxwell field of an IWP
solution satisfies
\bel{Max1n} X^\mu(F_{\mu\nu} + i*_4 F_{\mu \nu})dx^\nu =(F_{u\nu} + i*_4 F_{u \nu})dx^\nu = -dV=
O(r)dr+O(r^2)dy^a\;.
\ee
This shows that $\frac 1 \epsilon (F_{u\nu} + i*_4 F_{u \nu}) d\hat
x^\nu$ vanishes in the limit $\epsilon\to 0$. Also the term
$F_{ra}dr\wedge dy^a=\epsilon F_{ra}d\hat r\wedge d\hat y^a$, vanishes
in the limit, and it is now simple to check that the whole $F$
vanishes in the near-horizon limit. The near-horizon geometry of such
a solution is therefore a vacuum solution, with metric
\beq
{}\fourg =-2d\hat ud\hat r-\zh_{ab}d\hat y^ad\hat y^b.
\label{metric111} \eeq
Since $\fourg$ is Ricci flat, and $-2d\hat ud\hat r$ is flat, one obtains that
$\zh_{ab}d\hat y^ad\hat y^b$ is flat, contradicting the fact that the horizon must have
$S^2$ topology. Hence $Q$ cannot vanish identically.

The IWP metric (\ref{met1}) includes a flat 3-metric $d{\bfx} \cdot
d{\bfx}$. This is invariantly defined where $V$ is non-zero by
projecting the 4-metric orthogonally to $X$ and multiplying by
$V\overline{V}$. Calculating this in the Gaussian null coordinates
gives
\be
\label{flat}
 d{\bfx}\cdot d{\bfx} = \left( dr + r^2 H_a dy^a \right)^2 + r^2 A
h_{ab} dy^a dy^b.
\ee
The left hand side is flat so if we multiply the right hand side by
$1/\epsilon^2$ we should get another flat metric. Set $r=\epsilon r'$
in this new metric and now let $\epsilon \rightarrow 0$ to obtain the
metric
\be d{r'}^2 + 4 {r'}^2 |Q|^2 \zh_{ab} dy^a dy^b.
\ee
This must also be flat away from $r'=0$. By calculating the Ricci
scalar of this metric one finds that  $4 |Q|^2
\zh_{ab}$ has Ricci scalar equal to two, so it is the unit round
metric on $S^2$. But $\zh_{ab}$ is also a metric on $S^2$. Therefore
$Q$ cannot vanish anywhere.

Finally, consider the equation $\Delta V^{-1} = 0$. Writing this out
using the metric (\ref{flat}) we find that
\bel{Vlimeq}
 \tilde{\Delta} Q^{-1} = \lim_{r \rightarrow 0} \left( 8 r^3 |Q|^2
 \Delta V^{-1} \right) = 0,
\ee
where $\tilde{\Delta}$ is the Laplacian associated with the 2-metric
$\zh_{ab}$. Since $Q^{-1}$ is globally defined, this equation implies that
$Q$ must be constant. It then follows that $\zh_{ab}$ must be a
metric on $S^2$ of constant curvature $K=4|Q|^2=\zA$.\qed

\subsection{The supercovariantly constant spinors near the horizon}
\label{Sccs}

In this section, our aim is to obtain the supercovariantly constant
spinors near the horizon in order to relate the two coordinate
systems $(t, x^i)$ and $(u,r,y^a)$. Following \cite{CT}, in the metric
(\ref{metric1}), choose the coordinates $y^a$ so that they are
isothermal on ${\cal{N}}$ and then introduce
$\zeta=y^1+iy^2$. Choosing $m$ to be proportional to $d\bar \zeta$ at
$r=0$, the metric becomes
\be
\fourg
=r^2Adu^2-2dudr-2r^2(Hd\zeta+\overline{H}d\overline{\zeta})du-2m\overline{m}\;
\label{metric2} \ee
where
\[m=-\zX d\bar\zeta+O(r)\;,\]
in terms of a complex function $\zX $ of
$\zeta$ and $\overline{\zeta}$.

We shall investigate the metric (\ref{metric2}) in the
spin-coefficient formalism \cite{NewmanTod}. We introduce the null tetrad
$(l^\mu,n^\mu,m^\mu,\overline{m}^\mu)$ by
$$
\begin{array}{lllll}
\l^\mu\partial_\mu&=&D&=&\partial_u +\frac{r^2A}{2}\partial_r\;,\\
n^\mu\partial_\mu&=&\Delta&=&-\partial_r\;,\\
m^\mu\partial_\mu
&=&\delta&=&\frac{1}{\overline{Z}}\partial_{\zeta}
+\frac{r}{\overline{Y}}\partial_{\overline{\zeta}}
-\left(\frac{r^2H}{\overline{Z}}+\frac{r^3\overline{H}}{\overline{Y}}\right)\partial_r\;,
\end{array}
$$
where $Z=\zX +O(r)$.

We follow the numbering of \cite{NewmanTod} to calculate the spin-coefficients and curvature quantities. We may take the results from \cite{CT} by replacing $h$ there by $rH$, to give
\begin{eqnarray}
\label{SpinEq1}
\alpha&=&-\frac{1}{2\zX\overline{\zX}}\frac{\partial\overline{\zX}}{\partial\overline{\zeta}}+O(r)\;,\\
\beta&=&\frac{1}{2\zX\overline{\zX}}\frac{\partial \zX}{\partial\zeta}+O(r)\;,\\
\gamma&=&-\frac{1}{4}\frac{\partial}{\partial r}\log\left(\frac{\zX}{\overline{\zX}}\right)+O(r)\;,\\
\epsilon&=&\frac{1}{2}r\zA+O(r^2)\;,\\
\mu&=&-\frac{1}{2}\frac{\partial}{\partial r}\log(\zX\overline{\zX})+O(r)\;,
\label{SpinEq2}
\end{eqnarray}
together with $\pi = -\overline{\tau} =
O(r),\nu=0,\lambda=1+O(r),\rho=O(r^2), \kappa=O(r^2)$, and
$\sigma=O(r^2)$.  In Appendix~\ref{B} we give an alternative proof of
Proposition~\ref{Pgeom}, based on the above.

Expanding the spinor fields $\alpha_A$ and $\beta_{A'}$ in the spinor dyad as
\be
\alpha_A=-\alpha_0\iota_A+\alpha_1o_A\:;\quad \beta_{A'}=-\beta_{0'}\iota_{A'}+\beta_{1'}o_{A'}\;,
\label{comps1}
\ee
and substituting into (\ref{sc1}) and (\ref{sc2}), we obtain eight
equations from (\ref{comps1}) as follows (compare~\cite{Stewart:book} p.~219)
\begin{eqnarray}
D\alpha_0-\epsilon\alpha_0+\kappa\alpha_1+\sqrt{2}\phi_0\beta_{0'}&=&0\label{eq1}\\
D\alpha_1+\epsilon\alpha_1-\pi\alpha_0+\sqrt{2}\phi_1\beta_{0'}&=&0\label{eq2}\\
\delta\alpha_0-\beta\alpha_0+\sigma\alpha_1+\sqrt{2}\phi_0\beta_{1'}&=&0\label{eq3}\\
\delta\alpha_1+\beta\alpha_1-\mu\alpha_0+\sqrt{2}\phi_1\beta_{1'}&=&0\label{eq4}\\
\overline{\delta}\alpha_0-\alpha\alpha_0+\rho\alpha_1+\sqrt{2}\phi_1\beta_{0'}&=&0\label{eq5}\\
\overline{\delta}\alpha_1+\alpha\alpha_1-\lambda\alpha_0+\sqrt{2}\phi_2\beta_{0'}&=&0\label{eq6}\\
\Delta\alpha_0-\gamma\alpha_0+\tau\alpha_1+\sqrt{2}\phi_1\beta_{1'}&=&0\label{eq7}\\
\Delta\alpha_1+\gamma\alpha_1-\nu\alpha_0+\sqrt{2}\phi_2\beta_{1'}&=&0\label{eq8}
\end{eqnarray}
and the corresponding eight equations for $\beta_{A'}$, which can be
obtained from (\ref{eq1})-(\ref{eq8}) by complex conjugation followed
by the substitution of $-\alpha_A$ for $\overline{\beta}_A$ and
$\beta_{A'}$ for $\overline{\alpha}_{A'}$.

We also have two expressions for the Killing vector (\ref{Xdef}):
\begin{eqnarray*}
X^\mu&=&\frac{1}{\sqrt{2}}(\alpha^A\overline{\alpha}^{A'}+\overline{\beta}^A\beta^{A'})\\
&=& \delta^\mu_u \; =\; l^\mu+\frac{r^2A}{2}n^\mu\;.
\end{eqnarray*}
Substituting from (\ref{comps1}) into this, we obtain
\begin{eqnarray}
|\alpha_0|^2+|\beta_{0'}|^2&=&\frac{\sqrt{2}}{2}r^2A\;,\label{ab1}\\
\alpha_0\overline{\alpha}_{1'}+\beta_{1'}\overline{\beta}_0&=&0\;,\label{ab2}\\
|\alpha_1|^2+|\beta_{1'}|^2&=&\sqrt{2}\;.\label{ab3}
\end{eqnarray}
Thus, near the horizon, $\alpha_0$ and $\beta_{0'}$ are $O(r)$ while
$\alpha_1$ and $\beta_{1'}$ are $O(1)$. We write
\be
\alpha_1=\zZ+O(r)\;;\ \alpha_0=r\zS+O(r^2)\;;\ \beta_{1'}=\zW+O(r)\;
;\ \beta_{0'}=r\zT+O(r^2)\;,\label{asy}
\ee
and proceed to analyse the system (\ref{eq1})-(\ref{eq8}), using what
we know of the spin-coefficients and curvature components. From
Appendix~\ref{B} we have $\phi_0=O(r)$.
Equation (\ref{eq1}) is already $O(r^2)$; equations (\ref{eq2}),
(\ref{eq3}) and (\ref{eq5}) are $O(r)$; from the rest, (\ref{eq4}) and
(\ref{eq6}) and the corresponding equations for $\beta_{A'}$ yield the
system:
\begin{eqnarray}
(\delta+\beta)\alpha_1+\sqrt{2}\phi_1\beta_{1'}&=&O(r),\label{eq9}\\
(\overline{\delta}+\overline{\beta})\beta_{1'}-\sqrt{2}
\overline{\phi_1}\alpha_{1}&=&O(r),\label{eq10}\\
(\overline{\delta}+\alpha)\alpha_1&=&O(r),\label{eq11}\\
(\delta+\overline{\alpha})\beta_{1'}&=&O(r).\label{eq12}
\end{eqnarray}
Substituting from (\ref{asy}) into (\ref{eq11}) and (\ref{eq12}) we
obtain equations which can be readily solved to give
\be
\zZ=(\overline{\zX})^{\frac{1}{2}}f(\zeta)\;,
\quad
\zW=(\zX)^{\frac{1}{2}}g(\overline{\zeta})\;,\label{asy2}
\ee
for some holomorphic functions $f$ and $g$. It is convenient now to
take an explicit form for $\zX$. From Proposition \ref{Pgeom} we know
that $2\zA\zX\overline{\zX}d\zeta d\overline{\zeta}$ is the unit round
metric, thus if we introduce $\zL $ by $\zA=\zL ^{-2}$ then we may choose
\be
\zX=\overline{\zX}=\frac{\zL \sqrt{2}}{(1+\zeta\overline{\zeta})}.\label{exx}
\ee
The relation in Proposition \ref{Pgeom} $(iii)$ can now be written $4|Q|^2\zL ^2=1$, so that
\be 2Q\zL =e^{i\kappa}\label{Q} 
\ee
 for some real constant $\kappa$. Appendix~\ref{B} shows that
$\phi_1=Q+O(r)$.  With the choice \eq{exx} for $\zX$, we substitute
(\ref{asy2}) into (\ref{eq9}) and (\ref{eq10}) to obtain
\beaa
&
\zX^{-2}\partial_\zeta (\zX f) = - \sqrt 2 Q  g\;,
&
\\
&
\zX^{-2}\partial_{\bar\zeta} (\zX g) =  \sqrt 2 \bar Q  f\;.
&
\eeaa
Differentiating the first equation with respect to $\zeta$, and
expanding, one obtains an equation which can be solved for $f$;
inserting the result into the second, one is led to
\[f=a+b\zeta\;\;;\quad g=(a\overline{\zeta}-b)e^{-i\kappa}\;,
\]
for constant complex $a$ and $b$. We can exploit a residual
freedom in the choice of $\zeta$ to simplify these expressions.
Indeed, in view of the choice \eq{exx}, the $\zeta$'s are defined
now only up to a rigid $SO(3)$ rotation of $S^2$, which will
transform $a$ and $b$. Without loss of generality, we may assume
that $b=0$ and then, to satisfy (\ref{ab3}), that
$a=\zL ^{-\frac{1}{2}}$.

For $\zS$ and $\zT$ we go to (\ref{eq7}), and the corresponding
equation for $\beta_{0'}$, and solve to find
\[\zS=\sqrt{2}Q\zW\;\;;\quad \zT=-\sqrt{2}\overline{Q}\zZ.\]
Now we have the supercovariantly constant spinors to the desired
order, and may proceed to define the coordinates $x^i$ from
(\ref{xdef1}) and (\ref{xdef2}). This is a mechanical process, and we
simply record the results. Introduce polar angles via
\[\zeta=\tan(\theta/2)e^{-i\phi}\]
then
\begin{eqnarray}
dx^1+idx^2&=&d(-r\sin\theta e^{i\phi-i\kappa})+O(r)dr+O(r^2)(d\zeta, d\overline{\zeta}, du)\label{x1x2}\\
dx^3&=&d(-r\cos\theta)+O(r)dr+O(r^2)(d\zeta, d\overline{\zeta}, du).\label{x3}
\end{eqnarray}
If we introduce a set of Cartesian coordinates $z^i$ for $i=1,2,3$
related to the polar coordinates $(r,\theta,\phi)$ by
\be
z^1+iz^2=-r\sin\theta e^{i\phi-i\kappa}\;,\quad z^3=-r\cos\theta\;,
\label{xandz}
\ee
then the system (\ref{x1x2}) and (\ref{x3}) implies, for $z\ne 0$,
\bel{parder}
\frac{\partial x^i}{\partial z^j} = \delta^i_j +O(|z|)\;,
\ee
which will be important in the next section.

\subsection{Horizons are isolated singularities}
\label{Shis}

We equip $\hyp$ with the \emph{orbit space metric} $\gamma$ defined as
\begin{equation}
\gamma(Y,Z)=-{}\fourg (Y,Z)+ \frac{{}\fourg (X,Y)\;{}\fourg
(X,Z)}{{}\fourg (X,X)}\ ,
 \label{eq:hdefnew}
\end{equation}
where $X$ is the Killing vector field \eq{Xdef}.
Our main local result is the following:

\begin{Theorem}
\label{This} Every connected component of the horizon corresponds
to an isolated singular point\footnote{By ``isolated singular
point" we mean here a point in local coordinates as in \eq{met1},
with an isolated singularity of $V^{-1}$ there. We will see in
Section~\ref{Swmh} that the correct geometric interpretation is
that of ``cylindrical ends".} $\bfx _0$ of the orbit space metric
\eq{eq:hdefnew}. Furthermore, there exists $\rho>0$, a smooth
(perhaps $\C$-valued) harmonic function $\Uzero \in
C^\infty(B({\bfx _0},\rho))$, and real constants $m_0$, $n_0$ such
that in the small punctured coordinate ball $B^*({\bfx _0},\rho)$
near $\bfx _0$ we have
\beq \frac{1}{V}=\frac{m_0+in_0}{|{\bfx }-{\bfx_0}|} + \Uzero \;.
\label{Vforml} \eeq
\end{Theorem}

\begin{Remark}
{\em
This result provides an alternative proof of Proposition~2 of
\cite{ChNad}, which is the key step of the argument there.
}
\end{Remark}

\proof By the previous section there exists $\epsilon>0$ such that, in
the punctured $z$--cordinate ball $B^*(0,\epsilon)$, the following
holds
\bel{difeq}\frac{\partial x^i}{\partial z^j} = \delta^i_j +O(|z|)\;.
\ee
We want to show that the limits $\lim_{|z|\to 0} \bfx(z)$ exist.
In order to do that, consider any ray $[\lambda,\mu]\ni s\to
sz$, with $0<\lambda\le \mu<\epsilon$, by \eq{difeq} we have
 \bel{int1} \bfxk(\mu
z)-\bfxk(\lambda z)= \int_\mu^\lambda
\frac{d\left(\bfxk(sz)\right)}{ds}ds =(\mu-\lambda) z^k + O(|\mu
z|^2)\;.
 \ee
Let $\mu_n$ be any sequence converging to zero. Equation \eq{int1}
with $\mu=\mu_n$ and $\lambda=\mu_{n'}$ shows that $\bfxk(\mu_n
z)$ is Cauchy, therefore there exist numbers $\bfxk_0(z)$ such
that $\lim_{n\to\infty}\bfxk(\mu_n z)=\bfxk_0(z)$. We similarly
have, for any two points $z_1,z_2\in B^*(0,\epsilon)$, with
$z_1\ne -z_2$,
 \bel{int2} \bfxk(\mu_n
z_1)-\bfxk(\mu_n z_2)= \int_0^1 \frac{d\left(\bfxk(s\mu_n
z_1+(1-s)\mu_n z_2)\right)}{ds}ds = \mu_n(z^k_2-z^k_1) +
O(\mu_n^2)\;,
 \ee
 and passing to the limit $n \to \infty$ we obtain
 $$\bfxk_0(z_1)=\bfxk_0(z_2)\;.$$ Thus, the limits $\bfxk_0(z)$ are in
 fact $z$-independent, we will write $\bfxk_0$ for those limits from
 now on. Passing to the limit $n\to \infty$ in \eq{int1} with $\mu=1$
 and $\lambda=\mu_n$ we obtain now \bel{int3} \bfxk( z)-\bfxk_0= z^k +
 O(|z|^2)\;.  \ee This shows that $\lim_{|z|\to0} \bfxk(z)= \bfxk_0$,
 as claimed.  It also follows from this equation that the map
 \bel{ztox} z\mapsto x \ee is differentiable at the origin, with
 \eq{difeq} holding now both at the origin and in
 $B^*(0,\epsilon)$. Consequently, the map \eq{ztox} is continuously
 differentiable on $B(0,\epsilon)$.

The implicit function theorem shows that, decreasing $\epsilon$ if
necessary, the map \eq{ztox} is a diffeomorphism near $\bfx _0$.
Therefore $1/V$ is well-defined, as a function of $\bfx $, in a
small punctured ball $B^*({\bfx _0},\rho)$ near $\bfx _0$, and is
harmonic with respect to the Euclidean metric there by
\eq{harmeq}. By \eq{int3} we have $|z|=|{\bfx }-\bfx_0| +O(|{\bfx
}-\bfx_0|^2)$, and  \eq{Veq}
gives, away from $\bfx_0$,
$$\frac 1 V = \frac{1}{2Q|{\bfx }-\bfx_0|}
+O(1)\;.
$$
 Write $1/2Q= m_0+in_0$ and set $$\Uzero = \frac 1 V -
 \frac{m_0+in_0}{|{\bfx }-\bfx_0|}\;,$$ then $\Uzero $ is a bounded harmonic function on
 $B^*({\bfx _0},\rho)$.  By Serrin's removable singularity
 theorem~\cite[Theorem~1.19, p.~30]{Veronbook} $\Uzero $ can be extended
 through $\bfx _0$ to a smooth harmonic function on $B({\bfx _0},\rho)$.
\qed

For further reference we note the following: It follows from
Theorem~\ref{This} that near each event horizon the metric can be
written in the form \eq{met1}, with $V$ of the form \eq{Vforml}. From
\eq{metric2} we have
\bean
\fourg&=&r^2Adu^2-2dudr-2r^2{H_a}
dy^adu-h_{ab}dy^ady^b \\ \nonumber & = &
r^2A\Big(\underbrace{du-\frac{dr}{r^2A} -\frac{H_a}A
dy^a}_{dt+\omega_idx^i}\Big)^2 \\ && -\frac
1{r^2A}\Big(\underbrace{{dr^2} +{2 r^2 H_a} dy^a dr + {r^2}(A
h_{ab}+{r^2 H_aH_b})dy^ady^b}_{dx^idx^i} \Big)\;,
\label{metric2.n}
\eea
with $r^2 A = V \bar V$. The function $t$ is defined up to the
addition of a function $f=f(x^i)$, which corresponds to the ``gauge
transformation'' $\omega_i \to \omega_i +\partial_i f$.
The simplest choice for $t$ suggested by \eq{metric2.n} is
\bel{tchoice}
dt = du - \frac{dr} {r^2 \zA}\;.
\ee
An important consequence of \eq{metric2.n} is that $\omega_idx^i$ is a
well-defined one form in a neighbourhood of each connected component
of the horizon.

\section{Global arguments}\label{SGlobal}

\subsection{Asymptotics for large $r$}\label{Sas}

Consider an asymptotically flat end $\Sext$. The decay of the Maxwell
field and a standard analysis of \eq{sc1}-\eq{sc2} show that the
components of $\alpha$ and $\beta$ approach constant values as $|\bfx|$
tends to infinity in a $\delta$-parallel spin frame associated with the
flat euclidean metric $\delta$ on $\Sext$.

The global hypotheses of Theorem~\ref{T1n} show that the ADM
four-momentum of $\Sext$ is timelike. By~\cite[Section~3]{ChBeig1}
the Killing vector $X$ defined by \eq{Xdef} is strictly timelike,
$g(X,X)>\epsilon>0$. By boosting $\Sext$ within
the Killing development of $\hyp$ one can assume that $X$ is
asymptotically normal to $\Sext$, and a multiplicative
normalisation of the spinors $(\alpha,\beta)$ leads to
$X\to\partial_t$ as $r$ goes to infinity.

Again a straightforward asymptotic analysis of \eq{xdef1}-\eq{xdef2}
shows that, performing a rigid coordinate rotation if necessary, the
functions $x^i$ asymptote to the asymptotically flat coordinates of
\eq{falloff}, and in fact provide a coordinate system outside a large
compact set on $\Sext$, for $|x|$ large enough.

Now $V$, and therefore also $V^{-1}$, tends to a pure phase at
infinity, but $V^{-1}$ is harmonic so without loss of generality this
phase is constant. Again adjusting our choice of $\psi$ we may
suppose the phase to be zero. Now we have, for large $|x|$,
\beq
V=1+\frac{C}{|x|}+O(|x|^{-2})
\label{V}
\eeq
for a (complex) constant $C$. The usual asymptotic expansion of
stationary initial data~\cite{Simon:elvac} gives
\bel{omegadec}
\omega = O(|x|^{-2})\;,\quad \partial\omega = O(|x|^{-3}) \;.
\ee
Inserting \eq{V} into (\ref{curl}), we see that this decay of the
derivatives of $\omega$ is possible only if $C$ is real.

\subsection{Injectivity}
\label{Si}

The map which to a point of the space-time $(\mcM,{}\fourg )$
assigns the functions $(t,x^i)$ by solving the equations
\eq{xdef1}-\eq{xdef2} may fail to provide a global coordinate
system on $\mcM$. An example is provided by the usual maximal
extension of the degenerate Reissner-Nordstr\"om solutions: in
this space-time each point $(t,x^i)$ corresponds to an infinite
number of distinct points $p_{t,x}$ lying in distinct
asymptotically flat regions of $\mcM$. We emphasise that in this
example the functions $x^i$ are smooth and globally defined
throughout both $\mcM$ and $\mcM/\R$, where $\R$ is the action of
the flow of $X$, but the map which to a point in $\mcM/\R$ assigns
the coordinates $x$ fails to be injective.

Before proceeding further, recall that the Killing development
$(\mcM_{\mcK},g_{\mcK})$ of a hypersurface $(\hyp,{}\threeg )$
with Killing initial data $(N,Y)$ is defined~\cite{ChBeig1} as
$\R\times \hyp$ with the metric
\bel{Kdev}
g_{\mcK}:=N^2dt^2 -{}\threeg _{ij}(dx^i+Y^idt)(dx^j+Y^jdt)\;.
 \ee
 If a space-time $(\mcM,{}\fourg )$ contains $\hyp$ as a hypersurface with
unit normal $n$, and if we decompose a Killing vector $X$ as
$X=Nn+Y$, with $Y$ tangent to $\hyp$, then
$(\mcM_{\mcK},g_{\mcK})$ is isometrically diffeomorphic to the
subset of $\mcM$ obtained by moving $\hyp$ in the space-time
$(\mcM,{}\fourg )$ with the flow of $X$, when this flow is
complete, provided, e.g., that $X$ is causal and $\hyp$ is acausal
in $(\mcM,{}\fourg )$.

We have the following:
\begin{Theorem}
\label{Tglobal} Under the hypotheses of Theorem~\ref{T1n}, the
coordinate representation \eq{met1} on the Killing development of
$\hyp$ is global.
\end{Theorem}

\proof
Consider the manifold $\whhyp$ defined as follows: as a point set,
$\whhyp$ consists of the interior $\zhyp$ of $\hyp$ with an abstract
point $\bfxi$ added for each connected component of the event
horizon. The differentiable structure on~$\whhyp$ is the one induced
from $\zhyp$ away from the $\bfxi$'s, and the one coming from the $x$
coordinates as in Theorem~\ref{This} around each point
$\bfxi$.

On $\zhyp$,
considered as a subset of $\whhyp$, we introduce the metric
\bel{hgamdef}
\hgamm:=V\bar V \gamma\;,\ee
 where $\gamma$ is the orbit space metric \eq{eq:hdefnew}. The local
coordinate representation \eq{met1} shows that $\hgamm$ is flat. By
Theorem~\ref{This} the metric $\hgamma$ extends by continuity to a
smooth (flat) metric on $\whhyp$, still denoted by the same symbol
$\hgamma$.  Thus $(\whhyp,\hgamm)$ is a smooth, flat, Riemannian
manifold. By construction $\whhyp$ is the union of a compact set and
of an asymptotically flat region, and such manifolds are
complete\footnote{This can be established using, e.g., the arguments
of Appendix~B of~\cite{Chmass}.}. Again by construction, $\whhyp$ is
simply connected. By the Hadamard-Cartan theorem (see,
e.g.,~\cite{Lee:Rm}) $\whhyp$ is diffeomorphic to $\R^3$, with a
global manifestly flat coordinate system. This provides the global
coordinate representation \eq{met1}.  \qed

As a corollary we obtain

\begin{Corollary}
\label{Cgl}
Under the hypotheses of Theorem~\ref{T1n}, there exist constants $m_i,n_i\in \R$ such that
\bel{1Veq}
\frac{1}{V}=1+\sum_{j=1}^N\frac{m_j+in_j}{|{\bfx }-{\bfx_j}|}\;.
\label{Vform}
\ee
\end{Corollary}

\proof The hypotheses of Theorem~\ref{T1n} imply that $\hyp$ has a
finite numbr of boundary components, say $N$. By Theorems~\ref{This}
and \ref{Tglobal}, there exist constants $m_i,n_i\in \R$ such that the
function
$$U=\frac{1}{V}-1-\sum_{j=1}^N\frac{m_j+in_j}{|{\bfx }-{\bfx_j}|}$$
 approaches zero at infinity, and can be extended by continuity to a
smooth harmonic function on $\R^3$.  By the maximum principle $U\equiv
0$.  \qed

\subsection{Staticity}\label{Soa}

Suppose that $V$ is given by \eq{1Veq}.  We introduce $G$ and $\theta$
by
\be
V^{-1}=Ge^{i\theta}.
\label{phase}
\ee
Since $V^{-1}$ is nowhere zero, $\theta$ is well-defined. Since
$C$ in \eq{V} is real (see Section~\ref{Sas}) we have, for large
$|x|$, 
\begin{eqnarray}
G&=&1+\frac{M}{|x|}+O(|x|^{-2})\;,\label{ld1}\\
\theta&=&O(|x|^{-2})\;,\label{ld2}
\end{eqnarray}
where $M$ is the ADM mass of $\Sext$.

{}From (\ref{curl}) we now find
\bel{curleq}{\rm{curl\;}}\omega=2G^2\nabla\theta
\ee
which is divergence-free since $\chi$ is real.

 Set $U={\R }^3\setminus\bigcup_j B_j$, where $B_j$ is a small ball
around  ${\bfx_j}$.  By the divergence theorem (with signs
appropriately chosen)
\bel{dividentcoco}
\int_UG^2|\nabla\theta|^2dV=\oint_{S_{\infty}}G^2\theta\nabla\theta\cdot d{\bf{S}}
+\sum_{j=1}^N\oint_{S_j}G^2\theta\nabla\theta\cdot d{\bf{S}}.
\ee
Here $S_{\infty}$ is a sphere of large radius, and $S_j=\partial B_j$.
We shall show that the integrals at the right-hand-side vanish in the
obvious limit, proving vanishing of $\nabla \theta$.

As emphasised at the end of Section~\ref{Shis}, $\omega$ is a well
defined one-form near each component of the horizon, thus no ``Dirac
string'' singularities in the form discussed in~\cite{HartleHawking}
arise near the punctures.  Therefore any small topological two-sphere
$S$ around a puncture we have
\beq
\oint_S{\mathrm{curl\;}}\omega\cdot d{\bf{S}}=0\;.
\label{intcon}
\ee
Near
${\bfx_j}$ we have
\begin{eqnarray*}
G&=&\frac{(m_j^2+n_j^2)^{1/2}}{|{\bfx }-{\bfx_j}|}+O(1)\\
\sin\theta&=&\frac{n_j}{(m_j^2+n_j^2)^{1/2}}+O(|{\bfx
}-{\bfx_j}|)
\end{eqnarray*}
so that \eq{curleq} and (\ref{intcon}) entail
\begin{eqnarray*}
\oint_{S_j}G^2\theta\nabla\theta\cdot d{\bf{S}}&=&
\arcsin\left(\frac{n_j}{(m_j^2
+n_j^2)^{1/2}}\right)
\underbrace{\oint_{S_j}G^2\nabla\theta\cdot
d{\bf{S}}}_{=0}
\\
&&+\oint_{S_j}G^2O(|{\bfx }-{\bfx_j}|)\nabla\theta\cdot
d{\bf{S}}=O(|{\bfx }-{\bfx_j}|)\;,
\end{eqnarray*}
while,
\[\oint_{S_{\infty}}G^2\theta\nabla\theta\cdot d{\bf{S}}=O(|x|^{-3}).\]
It follows that the right-hand-side of \eq{dividentcoco} vanishes in
the limit as the $B_j$ shrink onto the ${\bfx_j}$ and $S_{\infty}$
recedes to infinity. Therefore so does the left and $\theta$ is
constant, but $\theta$ vanishes at infinity, so $\theta$ is everywhere
zero and $V$ is real. Thus the metric is a standard
Majumdar-Papapetrou metric. This completes the proof of
Theorem~\ref{T1n}.

\section{Solutions with maximal hypersurfaces}
\label{Swmh}

 As first pointed out by Sudarsky and
Wald~\cite{SudarskyWald92,SudarskyWald93}, maximal surfaces which are
Cauchy for the domain of outer communications provide a powerful tool
to study stationary black holes.  The existence of such hypersurfaces
in our context is an open question, which we will not address here,
but some comments are in order.

We want to calculate the mean extrinsic curvature, say $H_f$, of the
hypersurfaces $\{t=f(x^i)\}$.  For this, we note that, from \eq{metric2.n},
\beal{contmet} &\fourg^{\mu\nu}\partial_\mu \partial_\nu = ((r^2A)^{-1} -
r^2A|\omega|^2_\delta)\partial_t^2 + 2 r^2A \sum_i\omega_i \partial_i \partial_t -
r^2A \sum_i\partial_i \partial_i\;,\phantom{xxx}
&
\\
&
\sqrt{-\det \fourg_{\mu\nu}}= (r^2A)^{-1}\;,
\eeal{detg}
where $|\omega|^2_\delta = \sum_i\omega_i^2$, so that the field of
unit normals $n$ to the level sets of $\{t=f(x^i)\}$ takes the form
\bean &
\displaystyle
n_\mu dx^\mu = \frac{r\sqrt A}{\sqrt{1-r^4A^2|\omega+df|^2_\delta}}(dt-df)\;,
&
\\
& \displaystyle n^\mu\partial_\mu= n^0\partial_t+ \frac{r^3
A^{3/2}}{\sqrt{1-r^4A^2|\omega+df|^2_\delta}}
\sum_i(\omega_i+\partial_if) \partial_i \;, & \eeal{unitn}
where $n^0$ equals
$$
n^0=\frac{\Big(1-r^4A^2\sum_i\omega_i(\omega_i+
\partial_if)\Big)} {r\sqrt A \sqrt{1-r^4A^2|\omega+df|^2_\delta}}\;.
$$
This leads to
\bean
H_f&=& \nabla_\mu n^\mu =
\frac 1 {\sqrt {-\det \fourg_{\mu\nu}}} \partial_\alpha\Big(\sqrt {-\det \fourg_{\mu\nu}} n^\alpha\Big)
\\
&=&
r^2 A  \sum_i\partial_i\left(\frac{r\sqrt A
(\omega_i+\partial_if)}{\sqrt{1-r^4A^2|\omega+df|^2_\delta}}\right) \;.
\eeal{meanc}
The choice \eq{tchoice} leads to a one-form $\omega$ which, in the
coordinates $x^i$, satisfies near $x=0$
$$
\omega_i =O(|x|^{-1})\;,\quad \partial_j \omega_i =O(|x|^{-2})\;.
$$
 Equation~\eq{meanc} with $f=0$ leads to
$H_0= O(r)$, in particular $H_0$ vanishes in the limit $r\to 0$. A
rough estimate leads one to expect that there should exist a class
of solutions of the equation $H_f=0$ with the asymptotic behavior,
for small $r$,
 \bel{asf}
 f=O(\ln^2 r)\;.
 \ee
(This can probably be improved to $O(\ln r)$ by a more careful
inspection of the equation satisfied by $\omega$, but we have not
undertaken that analysis.) Solutions with such an asymptotic behavior
will lead to a geometry of the level sets of $f$ with decay rates
worse than \eq{wcbcid}-\eq{wbcid2} below, but more than sufficient for
\eq{RH} to hold.

Now, we expect that the arguments in~\cite{ChWald}, concerning the
global structure of globally hyperbolic domains of outer
communications, remain valid in the current setting, but we have not
checked this. Assuming this to be correct it is then standard,
using~\cite{bartnik:variational}, to construct an edgeless maximal
surface in $\mcM$, as a limit of a sequence of solutions of the
Dirichlet problem on compact sets.  In view of the analysis
in~\cite{Bartnik84} it is not completely unreasonable to expect
asymptotic flatness of $\hyp$, as well as ``weakly cylindrical''
behavior (as defined by \eq{RH} below) near the horizons, but this
remains to be proved. We are planning to return to this question in a
near future.

We continue with the analysis of the gravitational initial data near
 the horizons.  Let $n$ be the field of future directed unit normals
 to $\hyp=\{t=f\}$, and decompose the Killing vector field $X$ as
 $X=Nn+Y$, where $Y$ is tangent to $\hyp$. From \eq{unitn} we obtain
\bean
 \displaystyle X = \partial_t &= &
  \displaystyle
  \underbrace{\frac {r\sqrt A
\sqrt{1-r^4A^2|\omega+df|^2_\delta}}{\Big(1-r^4A^2\sum_i\omega_i(\omega_i+
\partial_if)\Big)}}_{N} n^\mu\partial_\mu
 \\
 &
 &
 \displaystyle  - \underbrace{\frac{r^4
A^{2}}{\Big(1-r^4A^2\sum_i\omega_i(\omega_i+
\partial_if)\Big)} \sum_i(\omega_i+\partial_if)}_{Y^i} \partial_i
\;.
 \eeal{unitn2}

 By inspection of \eq{metric2.n} and \eq{unitn2}, the gravitational
 initial data induced on $\{t=0\}$ behave, for small $r$, as
\beal{acbcid}
 &
 \threeg_{ij}=O_1(r^{-2})\;,\quad
\threeg^{ij}=O_1(r^{2})\;,
 &
 \\
 &
 N= O_1(r)\;,\quad  Y^{i}=O_1(r^3)\;, \quad Y_{i}=O_1(r)\;, \quad
K_{ij}=O(1)\;,
 &
\eeal{abcid2} where the equality $h=O_k(r^\sigma)$ means that
$\partial_{i_1}\cdots \partial_{i_\ell} f= O(r^{\sigma-\ell})$ for
$0\le \ell \le k$. Here the estimate on $K_{ij}$ can be obtained
from the equation $NK_{ij}=-\frac 12 \mcL_Yg_{ij}$.

To understand the geometry of the level sets of $t$ it is
convenient, near each puncture $x_i$, to return to spherical
coordinates centred at $x_i$, replacing $|x-x_i|$ by a new radial
coordinate $\rho$ defined as
$$
\rho= - \ln |x-x_i|\;.
$$
 We then obtain
\bel{cyl}
g:=\threeg_{ij}dx^i dx^j = A^{-1}d\rho^2 + \twog_{ab}dy^a dy^b\;,
 \ee
where the metrics
 $\twog_{ab}(\rho,\cdot)dy^a dy^b$ asymptote exponentially fast to the
 round metric on the sphere as $\rho$ tends to infinity. This shows
 that for $\rho$ large enough the space-metric is uniformly equivalent
 to a fixed, $\rho$--independent, product metric
 $$
 \mathring g:= d\rho^2 +\mathring g _{ab}dy^a
 dy^b
 \;.
 $$
Such metrics will be called \emph{weakly cylindrical}.

Using the variable $\rho$, \eq{abcid2} can be rewritten as
\beal{wcbcid}
 &
 N= O(e^{-\rho})\;,\quad  |dN|_{g}=O(e^{-\rho})\;,
 &
 \\
 &
 |Y|_{g}=O(e^{-2\rho}
 )\;,\quad |D
Y|_{g}=O(e^{-2\rho} )\;, \quad |K|_{g}=O(e^{-\rho})\;.
 &
 \eeal{wbcid2}
%
For the purposes of Theorem~\ref{Tmain2} we will need the following two hypotheses:
\beal{RH}
\begin{array}{l}
\mbox{1. the metric induced on $\hyp$ is weakly cylindrical, and }
\cr
\mbox{2. $|K|_g |Y|_g$ approaches zero as $\rho$ tends to infinity.}
\cr
\end{array}
\eea
The analysis above shows that this is clearly satisfied on the
level sets of $t$.

Next, consider the Maxwell field.   \Eq{dV} is equivalent to
 \bel{3decF}
  \nabla _\mu (\Re V) = F_{\mu\nu}X^\nu\;,\quad \nabla _\mu (\Im
V) = *_4F_{\mu\nu}X^\nu\;,
 \ee
where $*_4$ is the space-time Hodge dual.  As in Section~\ref{Sas}, we
choose $V$ so that $V\to 1$ in the asymptotically flat region, in
particular $\Im V$ approaches zero there.

  Let $E=E_idx^i$ be the electric field on $\hyp$ defined as $E_i =
F_{i\mu}n^\mu$, similarly let the magnetic field $B=B_idx^i$ on $\hyp$
be defined as $B_i = *_4 F_{i\mu}n^\mu$. (As usual, when performing
$3+1$ decompositions, the index $i$ refers to a coordinate systems so
that $\hyp=\{t=0\}$; this does not necessarily coincide with the
$t$-coordinate of \eq{met1}.) As already pointed out, we denote by
$\threeg$ the \emph{positive definite} metric induced by the
space-time metric $\fourg$ on $\hyp$ (so, in our current signature
convention, $\threeg_{ij}=-\fourg_{ij})$.  Furthermore, we let $D$
denote the covariant derivative operator associated with $\threeg$
(not to be confused with $D$ in Section~\ref{Snh}). Then
\beaa
&
B_i = \frac 12 \epsilon_i{^{jk}}F_{jk} \quad \Longleftrightarrow \quad
F_{ij}=\epsilon_{ijk}B^k\;,
&
\\
&
*_4F_{ij} = \epsilon_{ijk0}F^{k0}= \epsilon_{ijk} E^k\;,
\eeaa
 where $\epsilon_{ijk}$ is completely antisymmetric and equals
$\sqrt{\det \threeg}$ for $ijk=123$. Further, indices on three
dimensional objects are raised and lowered with $\threeg$.  From
\eq{3decF} we obtain
 \bel{3decF2}
  D _i (\Re V) = NE_i + F_{ij}Y^j\;,\quad D _i (\Im
V) = NB_i+*_4F_{ij}Y^j\;.
 \ee
Assume, now, that $\hyp$ is maximal. We then have the equations
$$
D_iK^i{_j}= 2F_{jk}E^k = 2\epsilon_{jk\ell}E^kB^\ell\;,\quad D_{(i}Y_{j)}=-NK_{ij}\;,
\quad D_i B^i=0\;,
$$
which together with the second equation in \eq{3decF2} lead to the
 divergence identity\footnote{The calculations here allow one to
 simplify considerably the arguments in
 \cite{SudarskyWald93,SudarskyWald92}. This will be discussed
 elsewhere.}
\bel{SWid}
D_i\Big(K^i{_j}Y^j -2(\Im V) B^i\Big) = -(|K|^2+2 |B|^2)N\;.
\ee
One integrates \eq{SWid} over a set which consists of $\hyp$ from
which coordinate balls $S_i(\epsilon)$ of radius $\epsilon$ around the
punctures have been removed. The boundary integral at infinity
vanishes by the asymptotic flatness conditions. Consider the integrals
$$
Q^B_i = \oint_{S_i(\epsilon)} B^i dS_i\;,
$$
 then $Q^B_i$ does not depend upon $\epsilon$, at least for $\epsilon$
small enough, since $B$ has vanishing divergence. This implies that
the boundary term involving $\Im V B^i$ gives a vanishing contribution
in the limit $\epsilon\to 0$ (recall that $V$, and hence also $\Im V$,
approaches zero at the punctures by Theorem~\ref{This}). Similarly,
the boundary contribution from the extrinsic curvature term vanishes
in the limit by \eq{wbcid2}. Hence
$$
\int_\hyp (|K|^2+2 |B|^2)N=0 \;.
$$
 Note that $N$ is strictly positive on the interior of $\hyp$ since
$X$ is causal, which shows that $K\equiv B\equiv0$. In particular the
Killing development of $\hyp$ is static. We now have $\mcL_Y g =0$, so
that $Y$ is a Killing vector of $g$ which approaches zero in the
asymptotic region. By~\cite{ChBeig1} $Y\equiv 0$, so $X=Nn$,  hence the
Killing vector is strictly timelike in the domain of outer
communications (understood as a subset of the Killing development) so
that Theorem~\ref{T1n} applies.

\section{Theorem~\protect\ref{T1n} and sources}
\label{Sadd}

In this section we relax the hypothesis that the space-time is
electrovacuum. It turns out to be straightforward to obtain a
version of Theorem~\ref{T1n} assuming instead that \emph{magnetic
currents vanish}, so that $\Im V^{-1}$ is harmonic. We thus
consider a space-time with a non-trivial super-covariantly
constant spinor field $\psi$, and assume again that there are no
null orbits of the associated Killing vector $X$.

Suppose, first, that there are no black holes. As already pointed out
in Section~\ref{sSIWP}, the (then globally defined) imaginary part of
$V^{-1}$ is harmonic in the (globally defined) metric $|V|^2\gamma$,
where $\gamma$ is the orbit space-metric \eq{eq:hdefnew}, hence $\Im V
=0$, then $\mbox{curl}\ \omega =0$ and the space-time is static. This
leads to a Majumdar-Papapetrou solution, either non-empty or flat.

Both the proof and the statement of Proposition~\ref{Pgeom} remain
unchanged. Indeed, the only difference is the need to analyse the
supplementary contribution $\chi X_\mu X_\nu$ to the energy-momentum
tensor. But $$\chi X\otimes X= \chi \partial_u \otimes \partial_u=
\epsilon^2 \chi \partial_{\hat u}\otimes \partial_{\hat
u}\to_{\epsilon\to0}0\;,$$ which shows that the near-horizon
space-time remains vacuum. The limit in \eq{Vlimeq} is not affected by
source fields which are smooth functions in the physical space-time,
and one concludes as before.

One finds by inspection that none of the equations of
Section~\ref{Sccs} is affected by a non-vanishing $\chi$.

The non-harmonicity of $\Re V^{-1}$ might result in a somewhat
different behavior of the function $\Re U_0$ appearing in
Theorem~\ref{This}, which now will be the sum of a constant and of a
contribution of the form $r\chi$ , for some function $\chi(r,y^a)$
which is smooth in spacetime (but not necessarily smooth with respect
to the coordinates $x^i$). Note that $\Im U_0$ remains smooth, as
before. This will lead to a correction $O(r)$ (with somewhat worse
differentiability: bounded derivatives, and second derivatives
$O(r^{-1})$) in the real part of $V^{-1}$ in \eq{Vform}, the imaginary
part of this last equation remaining of the same form as before.

The staticity argument in Section~\ref{Soa} applies without changes.
As a consequence, one is led to a Majumdar-Papapetrou space-time with
a potential $V$ such that $V^{-1}$ is a finite sum of monopoles and of a
bounded function.

  \appendix

\section{An alternative proof of Proposition~\ref{Pgeom}}
\label{B}

Here we give an alternative spinor proof of Proposition~\ref{Pgeom}.
>From \eq{SpinEq1}-\eq{SpinEq2}, we calculate the curvature components,
setting the scalar curvature to zero. For the Weyl spinor, we find
$\Psi_0=O(r^2), \Psi_1=O(r), \Psi_3=O(1)$, and $\Psi_4=O(1)$ together
with two expressions for $\Psi_2$. One forces $\Psi_2=O(r)$ while the
other is
\be
\Psi_2=\frac{1}{4}(A-K)+O(r)\label{psi2}
\ee
where $K=-\frac{1}{\zX \overline{\zX
}}\partial_{\zeta}\partial_{\overline{\zeta}}\left(\log(\zX
\overline{\zX})\right)$, which is the Gauss curvature of
${\cal{S}}$. Thus $A=K +O(r)$.

For the Ricci spinor, we find $\Phi_{00}=O(r^2), \Phi_{01}=O(r)$ and
the remaining components are $O(1)$. In particular we have
\beq
\Phi_{11}=\frac{1}{4}(A +K)+O(r).
\label{phi11}
\eeq
It follows from (\ref{phi1}) that $\Phi_{00}=2\phi_0\overline{\phi}_0$
and $\Phi_{11}=2\phi_1\overline{\phi}_1$. Thus the component $\phi_0$
of the Maxwell field is $O(r)$, while $\phi_1$ is $O(1)$; $\phi_1$ is
also constrained by the Maxwell equations, specifically by (A.5b) of
\cite{NewmanTod} which here becomes
\[\overline{\delta}\phi_1=O(r)\;,\]
or
\[\partial_{\overline{\zeta}}\phi_1=O(r)\;.\]
This integrates at once to give $\phi_1=Q+O(r)$ where $Q$ is
holomorphic in $\zeta$ on ${\cal{S}}$. It is also bounded (since it is
the contraction of the self-dual part of the Maxwell field with the
volume form of ${\cal{S}}$), and so it must be constant (the value of
this constant is proportional to the charge of the black hole). Now
from (\ref{psi2}) and (\ref{phi11})
\[\zA=K=4|Q|^2\]
which establishes $(i)$ and $(ii)$.

For $(iii)$ return to (\ref{dV}). Taking components along the null
tetrad, we find
\begin{eqnarray}\label{Veq1a}
\frac{\partial V}{\partial r}&=&2\phi_1\\\label{Veq1b}
\delta V&=&-2\phi_0\\\label{Veq1c}
\overline{\delta}V&=&r^2A\phi_2.
\end{eqnarray}
Thus $\delta V$ and $\overline{\delta}V$ vanish at ${\cal{S}}$ while
$\frac{\partial V}{\partial r}=2Q$, so that
\bel{Veq}V=2Qr+O(r^2).\ee
{}
\qed
\section{The mass-charge inequality}
\label{Acharge}

Consider the following spinor covariant derivative on $\hyp$:
 \bel{Sli1}
\nabla_i = D_i +A_i\;,\ee where $D_i$ is the standard spin
connection for spinor fields, and
\begin{equation}
  \label{Sli10}
  A_i= {{1\over 2 } K_{ij}\ga^j\ga_0}{- \frac 12
   {E^k}\ga_k\ga_i\ga_0}{- \frac 14
  \epsilon_{jk\ell}B^j\ga^k\ga^\ell\ga_i}\;.
\end{equation}
Here the $\gamma^\mu$'s are local sections of a bundle of
spinor-endomorphisms which, in an ON-frame for the Riemannian metric
$g$ on $\hyp$, satisfy the usual relation $\gamma^\mu \gamma^\nu
+\gamma^\nu\gamma^\mu = 2\diag(1,-1,-1,-1)$. By construction,
constancy of a spinor $\psi$ in this connection is the projection into
$\hyp$ of the equations (\ref{sc1}, \ref{sc2}).

The identity which lies at the heart of the mass-charge inequality
reads
 \bea\lefteqn{
D_i\langle\phi,(\nabla^i+\ga^i\ga^j\nabla_j)\phi\rangle = |\nabla
\phi|^2 -|\ga^i\nabla_i \phi|^2 } && \nonumber\\ && + \frac 14
\langle \phi,\Big\{\mu+ (\nu_i \ga^i+4\divE )\ga_0-4\divB
\ga^1\ga^2\ga^3 \Big\}\phi\rangle\;, \eeal{Sli24} where
\begin{eqnarray}\mu&=& R -|K|_g^2 +(\trg K)^2 - 2 |E|_g^2 -2
|B|_g^2\;,\label{Sli35.1} \\ \nu^j &=& 2 D_i (K^{ij} - \trg K
g^{ij}) -4\epsilon^j{}_{k\ell}E^kB^\ell\;. \label{Sli35.2}
 \end{eqnarray}
The matrix appearing in the second line of \eq{Sli24} is of the form
\bel{Adef}A:=a^\mu\ga_0\ga_\mu+b\ga_0+c\ga_1\ga_2\ga_3\;.
\ee
We note the inequality
\bel{sharp}
\langle \psi,A\psi\rangle\ge \left(
a^0-\sqrt{|\vec a|_\delta^2+b^2+c^2}\right) |\psi|^2\;,
\ee
with equality if and only if both sides vanish, in particular the
quadratic form $\langle \psi,A\psi\rangle$ is non-negative if and only
if
\bel{sharp2}
a^0\ge\sqrt{|\vec a|_\delta^2 +b^2+c^2}\;.
\ee
This shows that the second line of \eq{Sli24} will be non-negative
when \eq{pec}-\eq{pc2} hold. Furthermore, the vanishing of the
left-hand-side of \eq{sharp} with a non-trivial $\psi$ implies
equality in \eq{sharp2}.  Under the conditions of
Theorem~\ref{Tcharge}, standard arguments (see, e.g.,
\cite{BartnikChrusciel1,Herzlich:mass}) give existence of a solution
of the equation $\gamma^i\nabla_i\psi=0$ (with appropriate boundary
conditions~\cite{GHHP,Herzlich:mass} at a weakly trapped boundary, if
relevant), which in turn leads to the inequality \eq{mci}. The case of
equality leads to the existence of a spinor field $\psi$ satisfying
\bel{dpsi}
\nabla_i \psi = 0\;.
\ee
We set
\bel{kdata}
N=\langle \psi,\psi\rangle\;,\quad Y^i = \langle
\psi,\gamma^0\gamma^i \psi\rangle\;,
\ee
 and we consider the
Killing development $(\mcM_{\mcK},g_{\mcK})$ of $(\hyp,g,N,Y)$ as in
\eq{Kdev}. The electromagnetic field $F_{\mu\nu}$ can be constructed out of $E$ and $B$ on $\hyp$ in the obvious way, and we extend $F$ to $\mcM_{\mcK}$ by requiring $\mcL_X F=0$.

We need to show that $\psi$ extends to a super-covariantly constant
Killing spinor on $\mcM_{\mcK}$.

Splitting the (Dirac) spinor $\psi$ into a pair of 2-component spinors
$(\alpha_A, \beta_{A'})$, we write (\ref{dpsi}) as the projection into
${\cal{S}}$ of (\ref{sc1}, \ref{sc2}) as
\begin{eqnarray}
P_{\mu}{}^{\nu}(\nabla_{NN'}\alpha_B+\sqrt{2}\phi_{NB}\beta_{N'})&=&0\label{scp1}\\
P_{\mu}{}^{\nu}(\nabla_{NN'}\beta_{B'}-\sqrt{2}\overline{\phi}_{N'B'}\alpha_N)&=&0,\label{scp2}
\end{eqnarray}
where $P_{\mu}{}^{\nu}$ is the projection orthogonal to $n$ (thus
$P_{\mu\nu}={}^4g_{\mu\nu}-n_{\mu}n_{\nu}$).

Following (\ref{kdata}), the Killing vector $X=Nn+Y$ is given at
${\cal{S}}$ by
\bel{Xdef2} X=\frac{1}{\sqrt{2}} (\alpha^A\overline{\alpha}^{A'}
+\overline{\beta}^A\beta^{A'})\frac{\partial}{\partial x^{AA'}},
\ee
compare (\ref{Xdef}). We extend the spinors
$(\alpha_A, \beta_{A'})$ off of ${\cal{S}}$ by requiring their
Lie-derivative along $X$ to vanish. (Recall that the Lie-derivative of
a spinor field $\alpha_A$ along a Killing vector $X^a$ is defined as
\be {\mcL}_X\alpha_A:=X^{\mu}\nabla_{\mu}\alpha_A+\Phi_A{}^M\alpha_M
\label{LieS}
\ee
where the symmetric spinor $\Phi_{MN}$ is defined by
$$\nabla_{\mu}X_{\nu}=\Phi_{MN}\epsilon_{M'N'}+\overline{\Phi}_{M'N'}\epsilon_{MN}.
$$
See e.g. \cite[p.~40]{TodHuggett}).

Then (\ref{Xdef2}) holds throughout $\mcM_{\mcK}$. We define
$V=\alpha_A\overline{\beta}^A$ as in Section \ref{sSIWP} and then we
can calculate the derivative of $X$ at $\hyp$ and in directions
tangent to $\hyp$ from (\ref{scp1}) and (\ref{scp2}) as
\be
P_{\mu}{}^{\nu}\nabla_{\nu}X_{\beta}=P_{\mu}{}^{\nu}(\overline{V}\phi_{NB}\epsilon_{N'B'}
+V\overline{\phi}_{N'B'}\epsilon_{NB}).
\label{Xp}
\ee
However, in $\mcM_{\mcK}$, $X$ is a Killing vector so that
$\nabla_{(\mu}X_{\nu)}=0$. It follows that, at $\hyp$, we can omit the
$P_{\mu}{}^{\nu}$ in (\ref{Xp}). Then both sides have vanishing
Lie-derivative along $X$ and we recover equation (\ref{dX}) for the
derivative of $X$ throughout $\mcM_{\mcK}$.

By \eq{Xdef2} we have
$$
X^{AA'}\beta_{A'} = -\frac{\bar V}{\sqrt 2} \alpha^A\;,\quad
X^{AA'}\alpha_{A} = \frac{ V}{\sqrt 2} \beta^{A'}\;.
$$
We may use (\ref{LieS}) and the above to write the constancy of
$\alpha_A$ and $\beta_{A'}$ along $X$ in the form
\begin{eqnarray*}
X^{\mu}(\nabla_{MM'}\alpha_A +\sqrt{2}\phi_{AM}\beta_{M'})&=&0\\
X^{\mu}(\nabla_{MM'}\beta_{A'}-\sqrt{2}\overline{\phi}_{A'M'}\alpha_M)&=&0.
\end{eqnarray*}
Taken with (\ref{scp1}) and (\ref{scp2}), and since $X$ has a nonzero
component along $n$, this shows that (\ref{sc1}) and (\ref{sc2}) hold
at $\hyp$.

To complete the proof we need a result from spinor calculus: for a
Killing vector $X$ and any spinor field $\chi$
$$({\mcL}_X\nabla_{\mu}-\nabla_{\mu}{\mcL}_X)\chi=0.$$
To prove this, observe that it is true for any tensor in place of
$\chi$, so all that is necessary is to check it for the spinors
$\epsilon_{AB}$, $\epsilon_{A'B'}$; this property follows immediately
from \eq{LieS}.

Now all quantities in (\ref{sc1}) and (\ref{sc2}) have vanishing
Lie-derivative along $X$ so that, by virtue of holding at $\hyp$,
these equations hold throughout $\mcM_{\mcK}$: there is therefore a
supercovariantly constant spinor in $\mcM_{\mcK}$.

 The case $V=0$ leads to metrics which, locally, are
$pp$-waves~\cite{Tod,GibbonsHull} (not necessarily
electro-vacuum). Asymptotically flat $pp$-waves do not satisfy the
regularity hypotheses set forth here, except if the space-time metric
is flat~\cite{ChBeig1}. As flat metrics belong to the IWP family,
Theorem~\ref{Tcharge} is proved.

\bigskip

\noindent{\sc Acknowledgements:} We are grateful to the Isaac Newton
Institute, Cambridge, for hospitality and financial support. HSR is a
Royal Society University Research Fellow.

\bibliographystyle{amsplain}
\bibliography{%
../references/hip_bib,%
../references/reffile,%
../references/newbiblio,%
../references/newbiblio2,%
../references/bibl,%
../references/howard,%
../references/myGR,%
../references/newbib,%
../references/Energy,%
../references/netbiblio,%
../references/PDE}
\end{document}